\documentclass[twocolumn,aps,pra]{revtex4}
\usepackage{epsfig}
\usepackage[english]{babel}
\usepackage{latexsym}
\usepackage{graphics}
\usepackage{subfigure}
\usepackage{graphicx}
\usepackage{dcolumn}
\usepackage{amsmath}
\usepackage{hyperref}
\usepackage{amssymb}
\usepackage{color}

\begin{document}
\title{Response time of electron inside a molecule to light in strong-field ionization}

\author{J. Y. Che$^{1,\dag}$, Y. G. Peng$^{1,\dag}$, F. B. Zhang$^{1}$, X. J. Xie$^{2,*}$, G. G. Xin$^{3}$, and Y. J. Chen$^{1,\ddag}$}

\date{\today}

\begin{abstract}
We study ionization of aligned H$_2^+$ in strong elliptically-polarized laser fields numerically and analytically.
The calculated offset angle in photoelectron momentum distribution is several degrees
larger for the molecule than a model atom with similar ionization potential at diverse laser parameters.
Using a strong-field model that considers the properties of multi-center and single-center Coulomb potentials,
we are able to quantitatively reproduce this angle difference between the molecule and the atom.
Further analyses based on this model show that the response time of electron to light
which is encoded in the offset angle and is manifested as the time spent in tunneling ionization,
is about 15 attoseconds longer for the molecule than the atom.
This time difference is further enlarged when increasing the internuclear distance of the molecule.

\end{abstract}

\affiliation{1.College of Physics and Information Technology, Shaan'xi Normal University, Xi'an, China\\
2.College of Physics and Electronic Information, Luoyang Normal University, Luoyang, China\\
3.School of Physics, Northwest University, Xi'an, China}
\maketitle

\section{Introduction}
Photoelectric effect which is also called as photoemission is associated with the energy exchange between
these two  minimal  particles of electron and photon in the cosmos.
This effect therefore has its basic importance in both practice and theory \cite{Clauser,Adawi}.
It was first observed  in the light-induced electron emission from metal targets, and then explained
by Einstein's light quantum hypothesis, which greatly promoted the birth of quantum mechanics.
According to Einstein's theory,
this effect is triggered only when the light frequency (corresponding to photon energy) is larger than the ionization potential of
the valence electron in the metal target, and this triggering is unrelated to the light intensity.

With the development of laser technology, the light intensity has been significantly improved.
Now, it is known that this effect is general in the interaction between light and matter (including all solid, liquid and gas targets)  \cite{Pratt,Glover,Miaja,Sorokin}.
For gas cases, when the light intensity is high enough, the active electron inside an atom or molecule
can be emitted with the absorbtion of more than one photon \cite{Protopapas}.
Relevant phenomena have been termed as multi-photon ionization.
When the electric field of the laser related to the light intensity continues to increase
and is comparable to the field of the Coulomb potential of the target,
tunneling ionization is also triggered where the light intensity plays an important role in ionization of the active electron
from an atom or molecule  \cite{Yang1993,Becker2002}.

An open question in photoelectric effect is whether the photoemission process is instantaneous
or the response of the electron to light needs a finite time in the process \cite{Lawrence,Ossiander}.
Due to the limitations of experimental technique and quantum-mechanics principle,
this question has been not explored extensionally  until recently \cite{Krausz,Krausz2009,Maquet,Vrakking,Pazourek,Vos}.
For example, for single-photon ionization of Ar and He in weak high-frequency laser fields, experimental studies
have discovered a relative time delay of tens of attoseconds between ionization of valence and core electrons \cite{Schultze,Klunder},
This delay has been explained as the Eisenbud-Wigner time  \cite{Wigner}.
For tunneling ionization of He in strong low-frequency laser fields, a delay of about 100 attoseconds
related to the time the tunneling electron spends under the barrier has also been revealed in experiments \cite{Landsman}.
It is found that this delay agrees with the predictions of the Larmor time \cite{Buttiker} and
the time obtained with Feynman path integral approach \cite{Sokolovski,Yamada}, especially for the latter.

Very recently,  a semiclassical theory  is developed to describe the response time of the electron inside an atom
to light in strong-laser induced tunneling ionization \cite{Chen2021}. This theory defines the response time as
the time of the strong three-body interaction between electron, nucleus and photon.
The observables deduced from the response time predicted by this theory agree with recent attoclock experiments
for diverse laser and atomic parameters \cite{Eckle3,Boge,Torlina,Quan,Undurti}.
It therefore provides a new perspective for understanding temporal properties of strong-field tunneling ionization of atoms \cite{Eckle1,Eckle2,Eckle4}.
For molecules, one can anticipate that the interaction between the valence electron, the molecular multi-center Coulomb potential
and the laser electric field is different from the atomic single-center case.

In this paper, we focus on the response time of electron inside a molecule to a strong low-frequency laser field.
Through numerical solution of time-dependent Schr\"{o}dinger equation (TDSE), we study the ionization of aligned H$_2^+$, the simplest
diatomic molecular ion, in strong elliptically-polarized laser fields.
As changing the laser intensity and wavelength, the calculated offset angle in photoelectron momentum distribution (PMD)
is several degrees larger for the molecule than a model atom with similar ionization potential $I_p$.
This angle difference between the molecule and the atom is further enlarged when increasing the internuclear distance of the molecule.
With considering the multi-center characteristic of the molecular Coulomb potential in tunneling, we generalize the semiclassical
response-time theory  to the molecular case.
This generalized theory well reproduces the TDSE results for the molecule. 
It reveals that the response time of the molecule in strong-field tunneling ionization
is more than ten attoseconds longer than the atomic one.

\section{Theory methods}
\emph{TDSE}.-In the length gauge,
the Hamiltonian of the H$_2^+$ system interacting with a strong laser field can be written as (in atomic units of $ \hbar = e = m_{e} = 1$)
\begin{equation}
{H}(t)= H_{0}+\mathbf{E}(t)\cdot \mathbf{r}
\end{equation}
Here, $H_{0}={\mathbf{{p}}^2}/{2}+V(\mathbf{r})$ is the field-free Hamiltonian and
 $\text{V}(\textbf{r})$ is the Coulomb potential of the molecule.
This  potential $\text{V}(\textbf{r})$ used has the form of
$V(\mathbf{r})=-Z/\sqrt{\zeta+\mathbf{r}_1^{2}}-Z/\sqrt{\zeta+\mathbf{r}_2^{2}}$
with $\mathbf{r}_{1(2)}^2=(x\pm \frac{R}{2}\cos\theta')^2+(y\pm \frac{R}{2}\sin\theta')^2$
in two-dimensional cases. Here,
$\zeta=0.5$ is the smoothing parameter which is used to avoid the Coulomb singularity, and $\theta'$ is the alignment angle.
Here, we consider the case of the parallel alignment with  $\theta'=0^o$.
The term $Z=1$ is the effective nuclear charge. For the equilibrium separation of $R=2$ a.u., the ionization energy of the model molecule 
reproduced here is $I_p=1.11$ a.u.. For other internuclear distances of $R=1.8$ a.u. and $R=2.2$ a.u., the ionization energy reproduced 
is $I_p=1.146$ a.u. and $I_p=1.078$ a.u., respectively. 
To understand the role of multi-center Coulomb potential in tunneling, we compare ionization of H$_2^+$ to
a model atom with similar $I_p$. These parameters used in the expression of $\text{V}(\textbf{r})$
for the atomic case are $R=0$ and $Z=0.85$.

The term $\mathbf{E}(t)$ in Eq. (1) denotes the electric field of the laser.
In elliptically-polarized cases, the electric field $\mathbf{E}(t)$ used here has the  form of $\mathbf{E}(t)=f(t)[\vec{\mathbf{e}}_{x}E_{x}(t)+\vec{\mathbf{e}}_{y}E_{y}(t)]$, with $E_{x}(t)={E_0}\sin(\omega t)$
and $E_{y}(t)={E_1}\cos(\omega t)$, ${E_0}={E_L}/{\sqrt{1+\varepsilon^2}}$ and ${E_1}=\varepsilon{E_L}/{\sqrt{1+\varepsilon^2}}$.
Here, $E_L$ is the  maximal laser amplitude corresponding to the peak intensity $I$, $\varepsilon=0.87$ is the ellipticity,
$\omega$ is the laser frequency and $f(t)$ is the envelope function. The term $\vec{\mathbf{e}}_{x}$($\vec{\mathbf{e}}_{y}$)
is the unit vector along the $x(y)$ axis. We use trapezoidally shaped laser pulses with a total duration of fifteen cycles,
which are linearly turned on and off for three optical cycles, and then kept at a constant intensity for nine
additional cycles. The TDSE of $i\dot{\Psi}(\textbf{r},t)=$H$(t)\Psi(\textbf{r},t)$ is solved numerically
using the spectral method \cite{Feit} with a time step of $\triangle t=0.05$ a.u..
We have used a grid size of $L_x\times L_y=409.6\times 409.6$ a.u. with space steps
of $\triangle x=\triangle y=0.4$ a.u..
The numerical convergence is checked by using a finer grid.

In order to avoid the reflection of the electron wave packet from the boundary and obtain the momentum space wave function, the coordinate
space is split into the inner and the outer regions with
${\Psi}(\textbf{r},t)={\Psi}_{in}(\textbf{r},t)+{\Psi}_{out}(\textbf{r},t)$, by multiplication using a mask function
$F(\mathbf{r})$. The mask function has the form of
$F(\mathbf{r})=F(x,y)=\cos^{1/2}[\pi(r_b-r_f)/(L_r-2r_f)]$ for $r_b\geq r_f$ and $F(x,y)=1$  for $r_b< r_f$.
Here, $r_b=\sqrt{x^2+y^2/\epsilon^2}$, $r_f=2.1x_q$ with $x_q=E_0/\omega^2$  and $L_r/2=r_f+50$ a.u. with $L_r\leq L_x$.
In the inner region, the wave function ${\Psi}_{in}(\textbf{r},t)$ is propagated
with the complete Hamiltonian $H(t)$. In the outer region, the time evolution of the wave function ${\Psi}_{out}(\textbf{r},t)$ is carried out
in momentum space with the Hamiltonian of the free electron in the laser field.
The mask function is applied at each time  interval  of 0.5 a.u. and the obtained new fractions of the outer wave function are added to the momentum-space wave function $\tilde{{\Psi}}_{out}(\textbf{r},t)$ from which we obtain the PMD.
Then we find the local maxima of the PMD and the offset angle $\theta$ is obtained with a Gaussian fit of the angle distribution of local maxima.

\emph{TRCM}.-Let's start with the model developed in \cite{Chen2021} to describe the response time of electron inside an atom to light
in strong-field tunneling ionization. This model has been termed
as tunneling-response-classical-motion (TRCM) model. It arises from strong-field approximation (SFA)
\cite{Lewenstein1995} but considers the Coulomb effect \cite{MishaY,Goreslavski,yantm2010}.
This model first solves the SFA saddle-point equation
\begin{equation}
[\textbf{p}+\textbf{A}(t_s)]^2/2=-I_p
\end{equation}
to obtain the electron trajectory ($\textbf{p},t_0$).
Here, $\textbf{p}$ is the drift momentum of the photoelectron which does not include the Coulomb effect.
$\textbf{A}(t)$ is the vector potential of the electric field $ \mathbf{E}(t) $. $t_0$ denotes the tunneling-out time of the photoelectron.
It is the real part of the complex time $t_s=t_0+it_x$ that satisfies the saddle-point equation of Eq. (2).
The trajectory ($\textbf{p},t_0$) agrees with the following mapping relation
\begin{equation}
\mathbf{p}\equiv\mathbf{p}(t_0)=\textbf{v}(t_{0})-\textbf{A}(t_{0}).
\end{equation}
The term $\textbf{v}(t_{0})=\mathbf{p}+\textbf{A}(t_{0})$
denotes the exit velocity of the photoelectron at the exit position (i.e., the tunnel exit)  \cite{yantm2010}
\begin{equation}
\mathbf{r}_0\equiv\mathbf{r}(t_0)=Re(\int^{t_0}_{t_0+it_{x}}[\mathbf{p}+\mathbf{A}(t')]dt').
\end{equation}
The corresponding complex amplitude for the trajectory ($\textbf{p},t_0$)
can be expressed as $c(\textbf{p},t_0)\sim e^{b}$.
Here, $b$ is the imaginary part of the quasiclassical action
$S(\textbf{p},t_s)=\int_{t_s}\{{[\textbf{p}+\textbf{A}(t'})]^2/2+I_p\}dt'$ with $t_s=t_0+it_x$ \cite{Lewenstein1995}.

Then the TRCM assumes that at the tunnel exit $\mathbf{r}(t_0)$, the tunneling electron
with the drift momentum $\textbf{p}$  is still located at a quasi-bound state which approximately agrees with  the virial theorem.
A small period of time $\tau$ is needed for the tunneling electron to evolve
from the quasi-bound state into an ionized state. Then it is free
at the time $t_i=t_0+\tau$ with the Coulomb-included drift momentum $\textbf{p}'$.
This time $\tau$ can be understood as the response time of the electron inside an atom to light in laser-induced photoelectric effects and is manifested as the Coulomb-induced ionization time lag in strong-field ionization \cite{Xie,Wang2020}.
The mapping between the drift momentum $\textbf{p}'$ and the ionization time $t_i$ in TRCM is expressed as
\begin{equation}
\mathbf{p}'\equiv\mathbf{p}'(t_i)=\textbf{v}(t_{0})-\textbf{A}(t_{i}).
\end{equation}

The offset angle $\theta$ in PMD is related to the most probable route (MPR)
which corresponds to the momentum having the maximal amplitude in PMD. For MPR, the tunneling-out time $t_0$ of  the
photoelectron agrees with the peak time of the laser field. This angle $\theta$ satisfies the following relation
\begin{equation}
\tan\theta=p'_x/p'_y= A_x(t_i)/[A_y(t_i)-v_y(t_0)].
\end{equation}
The above expression has considered the factor that for the MPR $v_x(t_0)=0$.
By neglecting $v_y(t_0)$,  the adiabatic version of the above expression is also obtained. That is
\begin{equation}
\tan\theta\approx A_x(t_i)/A_y(t_i).
\end{equation}
The adiabatic version is applicable for $\gamma\ll1$. Here,  $\gamma=w\sqrt{2I_p}/E_0$
is the Keldysh parameter \cite{Keldysh}.
It has been termed as Coulomb-calibrated attoclock (CCAC)  \cite{Che2}.
In CCAC, with considering  $t_{i}=t_{0}+\tau $ and $\omega t_{0}=\pi/2$ for MPR,
we can further obtain the following relation
\begin{equation}
\tan\theta\approx\tan(\omega\tau)/\varepsilon\approx\theta\approx\omega\tau/\varepsilon
\end{equation}
for a small angle $\theta$. With these above expressions, when the lag $\tau=t_{i}-t_{0}$ is obtained analytically or numerically,
one can further obtain the offset angle $\theta$. In turn,
when the  angle $ \theta $ is obtained in experiments or TDSE simulations, one can also deduce the lag $\tau$ from this angle.

\emph{Atomic time Lag}.-According to  TRCM for atoms, at  $t_0$, the tunneling electron is still located
in a quasi-bound state $\psi_b$ with approximately agreeing with the virial theorem.
Specifically, the average potential energy of this state is
$\langle V(\mathbf{r})\rangle\approx V(\textbf{r}(t_0))$ and the average kinetic energy is
$\langle\textbf{v}^2/2\rangle=n_f\langle v_x^2/2\rangle\approx-V(\textbf{r}(t_0))/2$.
This state can be further approximately treated as a quasi particle
with the velocity $|\textbf{v}_{i}|=\sqrt{\langle v_x^2\rangle}\approx\sqrt{|V({\textbf{r}}(t_0))|/n_f}$
which is opposite to the direction of  the position vector $\textbf{r}(t_0)$ and points to the nucleus.
This  velocity reflects the basic symmetry requirement of the Coulomb potential on the electric state.
A time lag $\tau$ is needed for the tunneling electron to acquire an impulse from the laser field in order to break this symmetry.
Then for the MPR, the lag $\tau$ can be evaluated with the expression of
\begin{equation}
\tau\approx\sqrt{|V({\textbf{r}}(t_0))|/n_f}/E_0.
\end{equation}
Here, $n_{f}=2,3$ is the dimension of the system studied and the exit position
$\mathbf{r}_0\equiv\mathbf{r}(t_0)$ of Eq. (4) is determined by the saddle points of Eq. (2).
The form of $V(\textbf{r})$ used in actual calculations will be discussed later.
Once the value of $\tau$ is obtained,
by substituting $\tau$ into Eq. (6), we can obtain the TRCM prediction of the offset angle $\theta$.

\begin{figure}[t]
\begin{center}
\rotatebox{0}{\resizebox *{8.5cm}{6cm} {\includegraphics {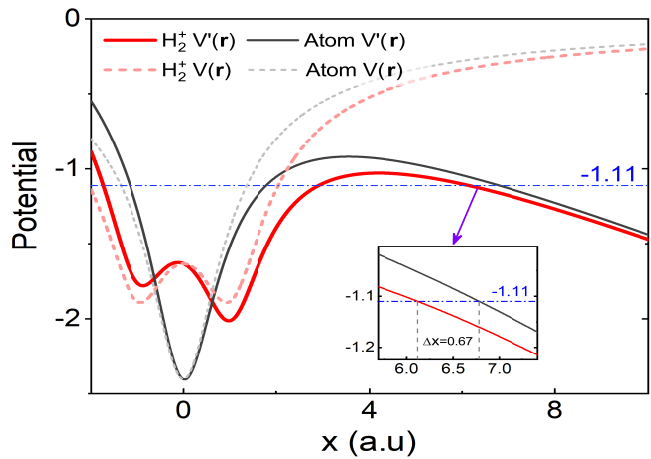}}}
\end{center}
\caption{A sketch of the laser-dressed potential $V'(\textbf{r})=V(\textbf{r})-E_0x$ (solid curves) and laser-free potential  $V(\textbf{r})$
(dashed-dotted) for H$_2^+$ (red curves) and the model atom (black) at $y=0$. The horizontal line indicates the energy of $-I_p=-1.11$ a.u..
The inset shows the enlarged results for the difference of exit position between the molecule and the atom.
The laser amplitude $E_0$ used here is $E_0=0.13$ a.u..
}
\label{fig:g1}
\end{figure}

\emph{Molecular time Lag}.-Next, we generalize the TRCM to molecules.
For the molecule with a relatively small internuclear distance, the high-energy bound states of the molecule
which have larger probability amplitudes far away from the nuclei, can be considered to be similar to the atomic ones and
approximately also agree with the virial theorem.
At the tunnel exit $r_0$, which is of the order of $r_0\sim 10$ a.u. for generally laser parameters used in experiments,
we assume that the tunneling electron of the molecule is still located in a quasi-bound state, which is consisted of high-energy
eigenstates of the field-free Hamiltonian H$_0$ of the molecule and approximately also agrees with the virial theorem.
To do so, we imply that the expression $E_0\tau=|\textbf{v}_{i}|\approx\sqrt{|V(\mathbf{r}(t_0))|/n_f}$ still holds for the molecule.
The main differences between the atom and the molecule are that they have different forms of Coulomb potential and therefore
1) the exit positions of the atom and the molecule differ from each other. As seen in Fig. 1,
the exit position  is about $0.7$ a.u. ($\sim R/2$) smaller for the molecule than the atom.
2) In addition, the left potential of the molecular two-center Coulomb potential is dressed up with a energy of $E_d=E_0R/2$.
This potential-dressed phenomenon around the nucleus disappears for the atom.
This phenomenon implies that the molecule with the dressed ionization potential $I'_p=I_p-E_d$ is somewhat easier to ionize than the atom.
The tunnel exit is of the order of $r_0\approx I_p/E_0$. With the above discussions of 1) and 2), we can assume that
the exit position of the molecule is $r_0'\approx I'_p/E_0-R/2\approx r_0-R$. Then we have
\begin{equation}
\tau\approx\sqrt{|V({\textbf{r}}'(t_0))|/n_f}/E_0.
\end{equation}
Here, ${r}'(t_0)\equiv r'_0=r_0-{R}$, with the value of $r_0$ evaluated by Eq. (4).
In single-active electron approximation, the potential  $V(\textbf{r})$ at the exit position $r_0$ can be considered to
has the form of $V(\textbf{r})\equiv V(r)=-Z'/r$. Here, $Z'$ is the whole effective charge.
For comparisons with TDSE simulations, the whole effective charge $Z'$ can be chosen as that used in simulations.
For comparisons with experiments, the value of $Z'$ can be evaluated with $Z'_a\approx\sqrt{2I_p}$ for atoms
and $Z'_m\approx I_p\sqrt{(R/2)^2+Z'^2_{a}/I_p^2}$ for molecules with the internuclear distance $R$. The expression of $Z'_m$
is obtained with assuming that $I_p\approx Z'_a/\sqrt{\zeta_0^2}$ for the companion atom
and $I_p\approx Z'_m/\sqrt{(R/2)^2+\zeta_0^2}$ for the molecule. Here, $\zeta_0$ is a constant factor.
This expression shows that with similar $I_p$, the value of $Z'$ is larger for molecules having larger $R$.
Because the exit position is smaller for the molecule than the atom
and the total effective charge $Z'_m$ of the molecule is larger than $Z'_a$ of the model atom,
the lag $\tau$ is also larger for the molecule than the atom.
Accordingly, considering $\theta\approx\omega\tau/\varepsilon$, the offset angle of the molecule is also larger than the atom.
In the following, we will show that the TRCM predictions of the molecular offset angle $\theta$,
evaluated by substituting Eq. (10) into Eq. (6), are in agreement with TDSE simulations.

\emph{PMDs}.-By assuming that for an arbitrary SFA electron trajectory ($\textbf{p},t_0$),
the Coulomb potential
 does not influence the corresponding complex amplitude  $c(\textbf{p},t_0)$,
we can obtain the TRCM amplitude  $c(\textbf{p}',t_i)$ for Coulomb-included electron trajectory ($\textbf{p}',t_i$) directly from the SFA one
with $c(\textbf{p}',t_i)\equiv c(\textbf{p},t_0)$ at $\tau\approx\sqrt{|V({\textbf{r}}(t_0))|/n_f}/|\textbf{E}(t_0)|$ for the atom and
$\tau\approx\sqrt{|V({\textbf{r}}'(t_0))|/n_f}/|\textbf{E}(t_0)|$ for the molecule.
This TRCM therefore allows the analytical evaluation of the Coulomb-included PMD.
The TRCM predictions of  PMD for H$_2^+$ and the model atom are shown in Fig. 2.

\begin{figure}[t]
\begin{center}
\rotatebox{0}{\resizebox *{8.5cm}{8cm} {\includegraphics {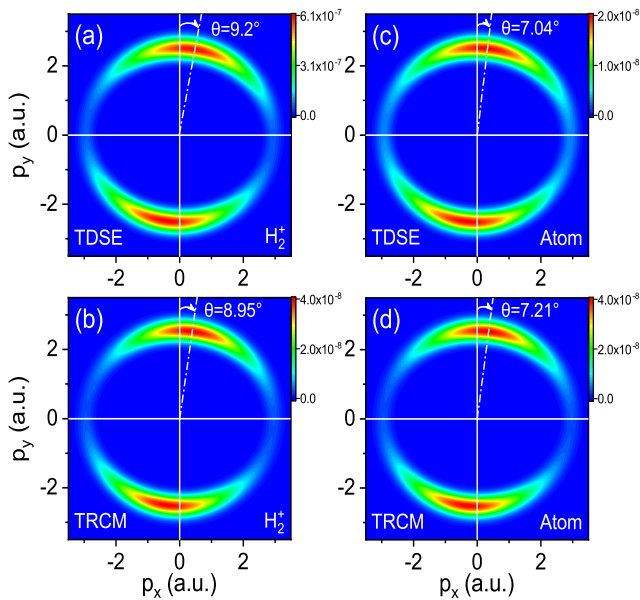}}}
\end{center}
\caption{PMDs of H$_2^+$ with $R=2$ a.u. (left column) and the model atom (right)  obtained with  TDSE (the first row)  and TRCM
(second) at  $I=1\times10^{15}$ W/cm$^{2}$ and $\lambda=1000$ nm.
The offset angle $\theta$ relating to the momentum with the maximal amplitude in PMD  is also indicated in each panel.
}
\label{fig:g2}
\end{figure}

\begin{figure}[t]
\begin{center}
\rotatebox{0}{\resizebox *{8.5cm}{8cm} {\includegraphics {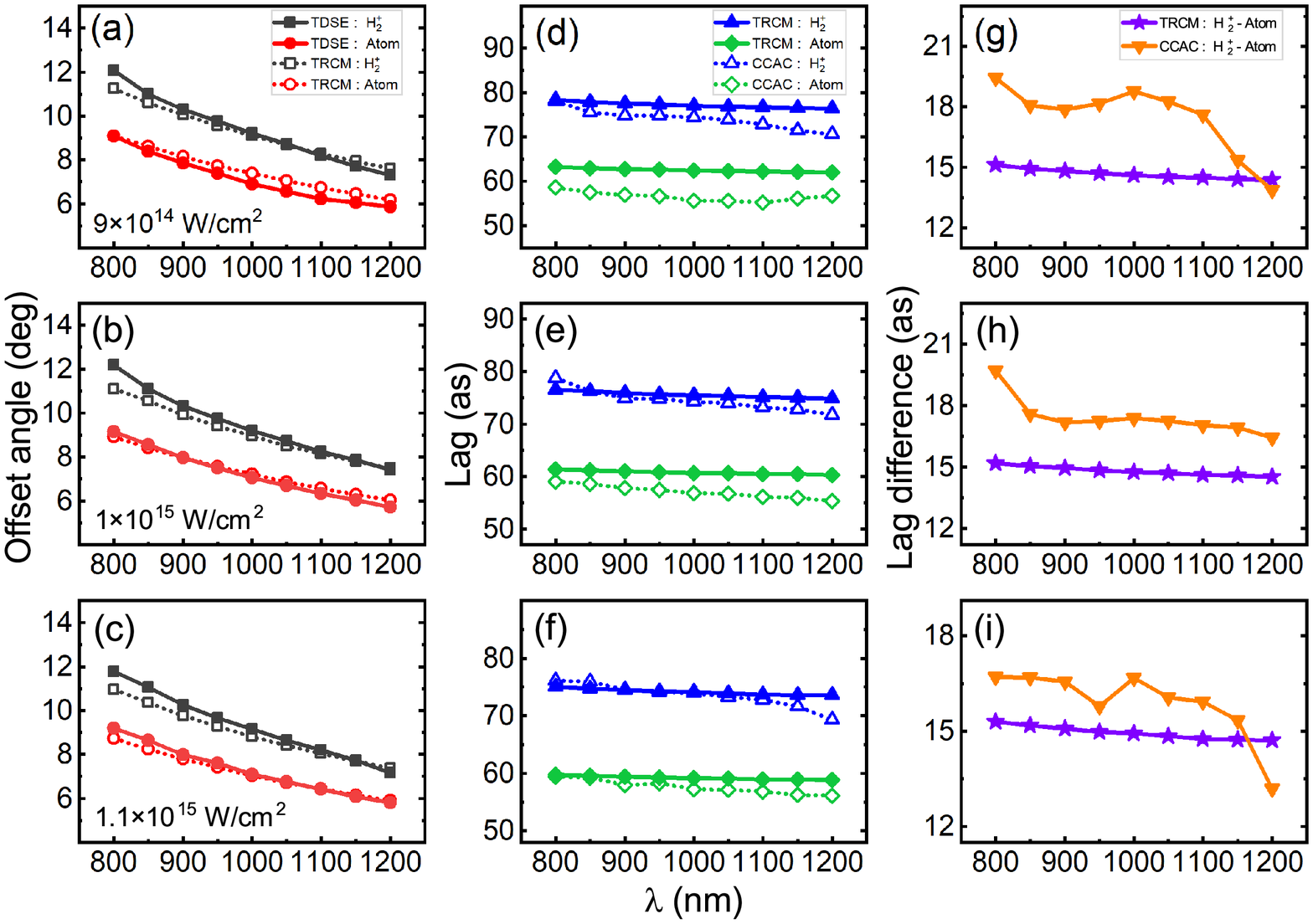}}}
\end{center}
\caption{Comparisons of the offset angle (the first column), the time lag $\tau$ (second) and the lag difference (third)
for H$_2^+$ with $R=2$ a.u. and the model atom  at $I=9\times10^{14}$ W/cm$^{2}$ (the first row),
$I=1\times10^{15}$ W/cm$^{2}$ (second) and $I=1.1\times10^{15}$ W/cm$^{2}$ (third) for different laser wavelengths $\lambda$.
The angles are obtained with TDSE and TRCM. The time lags and the corresponding lag differences are obtained with TRCM and CCAC.
In CCAC, the relation $\tau=\varepsilon\theta/\omega$ is used to evaluate this lag, with $\theta$ being the TDSE offset angle.
}
\label{fig:g3}
\end{figure}

\section{Results and discussions}
In Fig. 2, we show the PMDs of H$_2^+$ with $R=2$ a.u. and the model atom obtained with TDSE and TRCM.
First, the TDSE simulations show that the offset angle of the H$_2^+$ molecule in Fig. 2(a) is $\theta=9.2^o$,
while that of the model atom in Fig. 2(c) is $\theta=7.04^o$.
That is, with similar $I_p$, the offset angle of the molecule is about two degrees larger than the atomic one.
These TDSE results for the molecule and the atom are well reproduced by the molecular TRCM related to Eq. (10)
and the atomic one related to Eq. (9), respectively, as shown in Fig. 2(b) and 2(d).
The model predictions of the offset angle are $\theta=8.95^o$ for the molecule and $\theta=7.21^o$ for the atom,
very near to the TDSE results.
Our further analyses show that these three factors of the molecular larger effective charges $Z'_m$, the molecular laser-dressed energy $E_0R/2$
and the molecular nearer exit position contribute similarly to the increase of the offset angle of the molecule in comparison with the atom.
We mention that in this paper, we consider only the case where the molecular axis is aligned parallel to the major axis of laser polarization.
In this case, it has been shown that the molecular structure plays a small role in the offset angle \cite{Ren2022}.

To further explore these phenomena revealed in Fig. 2, we perform simulations at a wide range of laser parameters
and relevant results are shown in Fig. 3.
Firstly, one can observe from Fig. 3(a) that the TDSE offset angles of the molecule and the atom both decrease
with the increase of laser wavelength and the molecular offset angle is about 2 to 3 degrees larger than the corresponding atomic one
at different laser wavelengths.  These phenomena are well reproduced by the molecular TRCM and the atomic one.
When increasing the laser intensity, as seen in Figs. 3(b) and 3(c),  results are similar to Fig. 3(a), but the offset angle
at high intensity is somewhat smaller than the corresponding low-intensity one.

Next, we turn to the lag. In the second column of Fig. 3, we show the lag for the molecule and the atom calculated with Eq. (10) and Eq. (9),
respectively. For comparison, here, we also show the lag predicted by CCAC of $\tau=\varepsilon\theta/\omega$
with $\theta$ being the corresponding TDSE offset angle.
Firstly, one can observe from Fig. 3(d), the TRCM prediction of the lag with Eq. (10) for the molecule is about 88 attoseconds at different laser wavelengths and the lag predicted by Eq. (9) for the atom is about 63 attoseconds. The CCAC predictions of the lag
for the molecule and the atom are near to the corresponding TRCM ones with
a difference between TRCM and CCAC smaller than 10 attoseconds.
Results in Figs. 3(e) and 3(f) for higher laser intensities are similar to those shown in Fig. 3(d),
but the lag is several attoseconds smaller than the low-intensity one.

\begin{figure}[t]
\begin{center}
\rotatebox{0}{\resizebox *{8.5cm}{8cm} {\includegraphics {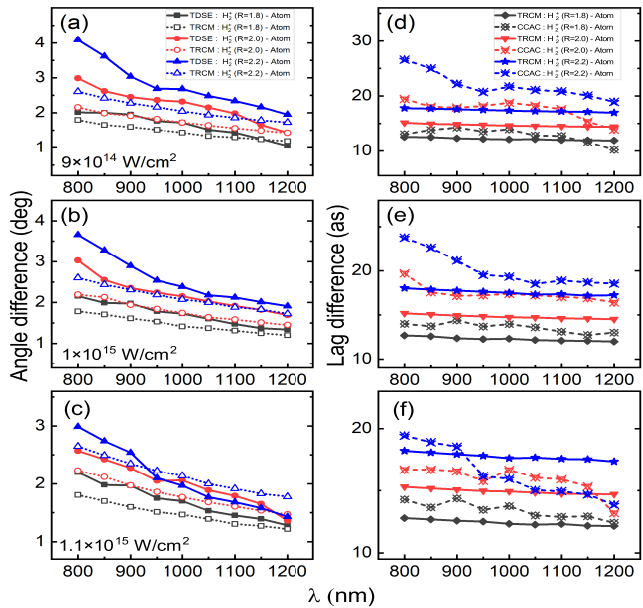}}}
\end{center}
\caption{Comparisons of the angle difference (the first column) and the lag difference (second)
between H$_2^+$ with diverse internuclear distances $R$ and the model atom at $I=9\times10^{14}$ W/cm$^{2}$ (the first row),
$I=1\times10^{15}$ W/cm$^{2}$ (second) and $I=1.1\times10^{15}$ W/cm$^{2}$ (third) for  different laser wavelengths $\lambda$. The angle differences are obtained with TDSE (solid curves) and TRCM (dotted) and
the lag differences are obtained with TRCM (solid curves) and CCAC (dashed).
}
\label{fig:g4}
\end{figure}

To highlight the difference of the lag between the molecule and the atom, in the third column of Fig. 3, we show the
lag difference for the corresponding results in the second column. One can observe that
the difference predicted by TRCM is about 15 attoseconds  and shows a slowly decreasing trend as the laser wavelength increases.
In addition, this difference seems insensitive to the laser intensity.
The CCAC predictions of the lag difference are similar to the TRCM ones, with the CCAC results being about 3 attoseconds larger than
the TRCM ones on the whole.

In Fig. 4, we further show the comparisons between H$_2^+$ with other internuclear distances of $R=1.8$ a.u.
and $R=2.2$ a.u. and the model atom at different laser intensities and wavelengths.
The results of $R=2$ a.u. presented in Fig. 3 are also plotted here as a benchmark.
Firstly, from the first column of Fig. 4, it can be seen that the difference of the offset angle
between the molecule and the atom obtained with TDSE
is mainly located at a range of about 1.5 degrees to 3 degrees, but for some short-wavelength cases of  $\lambda<900$ nm at $R=2.2$ a.u..
This TDSE angle difference is larger for the molecule having a larger $R$. In addition, for a fixed $R$, the angle difference shows a slowly
decreasing trend as the laser wavelength and the laser intensity increase.
The TDSE results are well reproduced by the molecular and the atomic TRCM related to Eq. (10) and Eq. (9), respectively.
The difference between TDSE and TRCM predictions is larger for cases of larger $R$ and shorter wavelengths.
It is smaller than 0.5 degrees on the whole.

The lag difference between the molecule and the atom predicted by TRCM and CCAC is shown in the second column of Fig. 4.
For the case of $R=1.8$ a.u., the lag difference of TRCM is not sensitive to laser intensity and wavelength,
and is around a value of about 12.5 attoseconds, smaller than 15 attoseconds of $R=2$ a.u..
For $R=2.2$ a.u., the TRCM difference is about 18 attoseconds at different laser parameters.
These results show the increasing trend of the lag difference with the increase of $R$, in agreement with the behavior of the angle difference.
The CCAC predictions of the lag difference are basically consistent with the TRCM ones,
but for the high-intensity case of $I=1.1\times10^{15}$ W/cm$^{2}$ with $R=2.2$ a.u. in Fig. 4(f).
For higher laser intensity, the ionization of H$_2^+$ with $R=2.2$ a.u. is also strong.
This effect is not considered in the present TRCM model.

We mention that since the ionization probability also plays an important role in the offset angle, and
a large ground-state depletion will remarkably decrease the offset angle \cite{Chen2021,Torlina}, in actual attoclock experiments for
a molecule and its companion atom, laser parameters related to applicable ionization yields for both targets are preferred.
In addition, in this paper, we compare the offset angle of H$_2^+$ with $1s\sigma_g$ symmetry to that of a model atom with $1s$ symmetry.
The symmetries of the molecule and the atom are near in our simulations.
In actual experiments, the symmetries of a molecule and an atom with similar $I_p$ may differ remarkably
from each other. This symmetry difference can also play a role in the comparison of the offset angle between the molecule and the atom.
Near symmetries for the molecular and the atomic targets are preferred in such a comparison.

\section{Conclusion}
In summary, we have studied ionization of aligned H$_2^+$ with different internuclear distances $R$
in strong elliptically-polarized laser fields, with comparing to a reference atom with similar ionization potential $I_p$.
Our TDSE simulations showed that the offset angle in PMD of H$_2^+$ is several degrees larger than the atomic one and
the angle difference between the molecule and the atom increases with increasing $R$.
By using a developed strong-field model termed as TRCM which considers the two-center and single-center characteristics
of the molecular and atomic Coulomb potentials, we are able to quantitatively reproduce these TDSE results.
This model reveals that with different forms of Coulomb potential but similar $I_p$,
the electron inside a molecule and that inside an atom response differently to a strong-laser induced ionization event.
During the process of ionization through tunneling, the barrier formed by the laser field and the molecular two-center Coulomb potential
is lower and narrower than the atomic single-center one, but the tunneling electron of the molecule ``feels" a stronger Coulomb force.
This is due to that for the molecular case, the tunnel exit  is nearer to the nuclei and the whole effective charge is also larger.
Compared with the atomic case, in order to overcome the stronger Coulomb potential remaining at the tunnel exit,
the tunnel electron of the molecule has to spend a longer time to obtain the enough impulse from the laser field.
This time has been termed as the response time of electron inside a molecule or an atom to light in strong-field tunneling ionization.
For cases studied in the paper, the molecular response time is about 10 to 20 attoseconds larger than the atomic one.
Our results shed light on attosecond-resolved studies of tunneling dynamics of molecules in strong laser fields.

This work was supported by the National Natural Science Foundation of China (Grant No. 12174239),
and the Fundamental Research Funds for the Central Universities of China (Grant No. 2021TS089).

\end{document}